%% file: main.tex
\documentclass[aps,prd,twocolumn,floatfix,preprintnumbers,superscriptaddress,nofootinbib,nameinlink,capitalise]{revtex4-2}
\usepackage{style, slashed}
\usepackage[dvipsnames]{xcolor}

\begin{document}

\title{$\mathbf{\mu}$-hybrid inflation and metastable cosmic strings in $\mathbf{SU(3)_c\times SU(2)_L\times SU(2)_R \times U(1)_{B-L}}$}
\input{authors.tex}
\noaffiliation

\begin{abstract}
We develop a scenario based on $\mu$-hybrid inflation within the framework of a left-right symmetric model defined by the gauge symmetry group $G_{3221} = SU(3)_c \times SU(2)_L \times SU(2)_R \times U(1)_{B-L}$. This model features a two-stage symmetry breaking to the Standard Model, with inflation occurring between the two stages, resulting in the dilution of primordial monopoles produced during the symmetry breaking of $SU(2)_R$ to $U(1)_R$. The second symmetry breaking gives rise to a network of metastable cosmic strings, which produces gravitational waves that can be observed by current and future experiments. A crucial aspect of this model is the inclusion of $R$-symmetry breaking terms in the superpotential, which helps explain the origin of the first symmetry breaking scale and ensures experimentally viable inflation with a minimal (canonical) K\"ahler potential. We show that the model successfully implements consistent inflation, leptogenesis, metastable cosmic strings, and gravitino dark matter, all in agreement with the observational data and constraints on reheating temperature from leptogenesis and gravitino overproduction bounds. We explore the potential embedding of $G_{3221}$ within the gauge symmetry $G_{422} = SU(4)_c \times SU(2)_L \times SU(2)_R$, and provide a brief discussion on the proton decay predictions, which can be tested by the upcoming experiments.
\end{abstract}

\maketitle

\newpage
\section{Introduction} \label{intro}
The supersymmetric (SUSY) hybrid inflation scenario naturally associates inflation with symmetry breaking, usually based on grand unified theories (GUTs). This framework utilizes a unique renormalizable superpotential consistent with gauge and $R$-symmetries, supported by inflation-friendly radiative and other crucial corrections \cite{Dvali:1994ms, Copeland:1994vg,Linde:1997sj,Buchmuller:2000zm,Senoguz:2004vu,Rehman:2009nq,Bastero-Gil:2006zpr,urRehman:2006hu,Rehman:2009yj,Civiletti:2014bca}. These corrections guide the inflaton toward the the minimum of the potential, ultimately ending inflation through a waterfall transition. The predictions of the model align with the latest cosmic microwave background (CMB) observations from Planck 2018. Additionally, the potential for observable primordial gravitational waves in supersymmetric hybrid inflation has been highlighted in Refs. \cite{Shafi:2010jr,Rehman:2010wm,Ahmed:2024rdd,Rehman:2018nsn} in the context of B-mode polarization experiments \cite{Ade_2019, Abazajian_2022, Belkner_2024, 10.1093/ptep/ptac150, Aurlien_2023, hanany2019picoprobeinflationcosmic, Finelli_2018, A_Kogut_2011,Andr__2014}.

In the framework of SUSY hybrid inflation, the term ``$\mu$-hybrid inflation'' \cite{Dvali:1997uq,King:1997ia,Okada:2015vka,Wu:2016fzp,Rehman:2017gkm,Afzal:2022vjx,ahmed2024inflation} refers to a specific class of models that, among other advantages, addresses the well-known $\mu$-problem of the minimal supersymmetric standard model (MSSM). 
This issue is resolved by introducing an additional trilinear coupling $S H_u H_d$ in the superpotential. This coupling generates the desired MSSM $\mu$-term through a nonzero vacuum expectation value (VEV), $\langle S \rangle \propto m_{3/2} $, induced by soft SUSY breaking terms involving the gravitino mass $m_{3/2}$ \cite{Dvali:1997uq,King:1997ia}. In addition, it also provides a dominant decay channel for the inflaton and contributes essential radiative corrections, ensuring that the predictions of the model align with the observational data. For Higgs inflation models based on the $\mu$-hybrid inflation framework within no-scale supergravity, utilizing a logarithmic form of the K\"ahler potential, refer to Refs. \cite{Okada:2017rbf,Ahmed:2021dvo}. For the shifted and smooth variants of $\mu$-hybrid inflation, refer to Refs. \cite{Lazarides_2021,Afzal:2023cyp,Zubair:2024quc}.
In the standard version of $\mu$-hybrid inflation scenario, any topological defects related to the breaking of the underlying symmetry emerge at the end of inflation. For instance, in certain cases, the gauge symmetry breaking patterns that eventually give rise to the Standard Model (SM) can lead to the formation of metastable cosmic strings --- the central focus of this study --- which later on decay into gravitational radiation.

A recent analysis of the 15-year pulsar timing data from the North American Nanohertz Observatory for Gravitational Waves (NANOGrav) has provided strong evidence for a stochastic gravitational wave background (SGWB) signal \cite{Agazie_2023}. One compelling interpretation of this signal arises from metastable cosmic string networks \cite{TWBKibble_1976, VILENKIN1982240}, which are well supported by models of physics beyond the Standard Model. In particular, SUSY models have gained significant attention for offering an elegant framework to explain the existence of metastable cosmic strings \cite{antusch2023singling, antusch2024explaining, buchmuller2024metastable, Fu_2024, chitose2024, Ahmed:2023rky, Afzal:2023cyp, maji2023supersymmetric, ahmed2023gravitational, King_2024, ahmed2024inflation}. A recent study in Ref. \cite{antusch2024probing} takes a slightly different approach, highlighting the potential for future gravitational wave observatories to uncover evidence of SUSY. The detection of SGWB from metastable cosmic strings is promising not only for pulsar timing arrays but also for interferometric detectors like the Laser Interferometer Gravitational-Wave Observatory (LIGO), the Einstein Telescope, Cosmic Explorer, and upcoming space-based observatory, the Laser Interferometer Space Antenna (LISA).

This study focuses on investigating a minimal $\mu$-hybrid inflation scenario based on the left-right gauge symmetry group, denoted as $G_{3221}\equiv SU(3)_c\times SU(2)_L\times SU(2)_R\times U(1)_{B-L}$. The left-right symmetric models offer several attractive features, such as quark-lepton unification, the natural emergence of right-handed neutrinos (RHNs), the generation of tiny neutrino masses via the seesaw mechanism, and an explanation for the baryon asymmetry \cite{PhysRevD.10.275, PhysRevD.11.2558, PhysRevD.12.1502, Senjanovic1979334, Dar2006517, Cvetivc1984n, PhysRevD.48.4352, PhysRevLett.75.3989, PhysRevLett.79.2188, PhysRevD.57.4174}. Of late, these models have been extended to encompass concepts like inflation and dark matter (e.g., Refs. \cite{Dvali:1997uq, Khalil_2012, Maleknejad:2020pec, Maleknejad:2020yys}). Notably, $\mu$-hybrid inflation was first introduced within the left-right symmetric configuration in Ref. \cite{Dvali:1997uq}, in which inflation was explored by considering radiative corrections alone.
This study also accounts for other important contributions, including those arising from supergravity and the soft SUSY breaking terms. Additionally, we incorporate $R'$-symmetry breaking\footnote{$R'$ corresponds to $R$ of $U(1)_R$ global symmetry, which hereinafter is written as such to distinguish it from the $R$ of $SU(2)_R$ or $U(1)_R$ gauge symmetries.} (denoted as $\slashed{R'}$) terms in the superpotential at the nonrenormalizable level, both to enhance the phenomenological aspects of the model and to provide an explanation for the origin of the first symmetry breaking scale.
We aim to achieve successful inflation that coherently relates to subsequent processes like reheating, leptogenesis, and formation of metastable cosmic strings while ensuring compatibility with the latest experimental data.

The proposed scheme for this study involves a two-stage breaking of symmetries, starting with an assumed initial symmetry breaking of $SU(2)_R \rightarrow U(1)_R$. Observable inflation is supposed to occur following this first breaking, driven by a gauge singlet inflaton coupled to a pair of $SU(2)_R$ doublet waterfall Higgs fields and a pair of electroweak Higgs doublets arising from the $SU(2)_L \times SU(2)_R$ symmetry. The monopoles generated during the initial symmetry breaking are subsequently inflated away. After inflation, the inflaton decays into RHN, reheating the Universe and facilitating baryogenesis via primordial leptogenesis. A direct coupling between the inflaton and the electroweak Higgs doublets provides a second, more dominant decay channel, producing a pair of Higgsinos \cite{Lazarides1998}. Following the inflationary phase, a second symmetry breaking occurs as $U(1)_R \times U(1)_{B-L}$ transitions into $U(1)_Y$, resulting in the formation of cosmic strings. 
This cosmic string network is metastable, capable of decaying in the presence of monopole-antimonopole pairs produced through the Schwinger process \cite{Buchmuller2020135764}, leading to the generation of a SGWB signal within the detectable range of gravitational wave experiments. 

Recently, the generation of metastable cosmic strings from a symmetry breaking pattern similar to that of the present model, has been investigated in Refs. \cite{Buchmuller:2021dtt, buchmuller2024metastable}, incorporating standard hybrid inflation. In contrast, the present study employs the $\mu$-hybrid inflation framework while accounting for the effects of $R'$-symmetry breaking terms in the superpotential. Furthermore, we provide details on mechanism of the first symmetry breaking and origin of its scale, as detailed in \cref{GUTorigin}.  A more recent work presented in Ref. \cite{Pallis:2024joc} investigates the formation of such strings in the context of T-model Higgs inflation, driven by an $SU(2)_R$ triplet superfield, utilizing a distinct choice of the superpotential and K\"ahler potentials.

A supplementary part of this study is to explore the possibility of proton decay through the potential embedding of $G_{3221}$ group into the larger gauge group $G_{422}\equiv SU(4)_c \times SU(2)_L\times SU(2)_R$ \cite{PhysRevD.10.275}. However, this necessitates the addition of a few extra fields in the content of $G_{3221}$, which can explain the proton decay predictions of $G_{422}$, primarily through the contribution of chirality nonflipping LLRR-type (where LLRR refers to the chirality of the operator, with ``L" and ``R" denoting, respectively, left-handed and right-handed chirality) proton decay operators. These predictions are expected to be observable in the next-generation proton decay experiments, such as Hyper Kamiokande \cite{Hyper-Kamiokande:2018ofw} and DUNE \cite{DUNE:2020ypp,DUNE:2015lol}.

Here is a brief outline of what follows. \Cref{model} describes the phenomenology of the $G_{3221}$ model and summarizes its basic features. A comprehensive discussion on inflation is provided in \cref{inflation}. It describes the scalar potential using minimal K\"{a}hler potential and the motivation for incorporation of the various terms in it. The process of reheating along with bounds on the reheat temperature is described in \cref{reheating}. \Cref{strings} briefly elaborates on the production and decay of metastable cosmic strings. A short discussion to explore the viability of gravitino as cold dark matter (DM) candidate within the parameter space of this model is covered in \cref{dm}. Numerical results obtained within the scope of the proposed model are discussed in \cref{numerics} along with the assumptions and constraints considered for that purpose. The relevant details on leptogenesis are in \cref{leptogenesis}. \Cref{422emb} discusses the possible incorporation of $G_{3221}$ into $G_{422}$, providing a succinct explanation of proton decay predictions within the $G_{422}$ configuration. We summarize the key aspects of the study and conclude in \cref{con}.
\section{Supersymmetric \texorpdfstring{$\mathbf{SU(3)_c \times SU(2)_L \times SU(2)_R \times U(1)_{B-L}}$}{3221} Model}\label{model}

\begin{table}[t!]
\caption{\label{content} Notations used below represent gauge groups as $G_{3221}=SU(3)_c\times SU(2)_L\times SU(2)_R\times U(1)_{B-L}$ and $G_{\text{SM}}=SU(3)_c\times SU(2)_L\times U(1)_{Y}$. Note that $p$ is an integer, the value of which is fixed $\geq 3$. The values of $p$ and $R'$-charge $q$ of $\Delta$ are further discussed in \cref{model}.}
        \begin{ruledtabular}
        \begin{tabular}{ccccc}
             &
            $ \mathbf{G_{3221}}$& $\mathbf{q(U(1)_{R'})}$ &
            $ \mathbf{G_{\text{SM}}}$&\\
                \hline
            &$Q(3,2,1,1/6)$ & $1/2$ & $Q(3,2,1/6)$&\\
            & $Q^c(\bar{3},1,2,-1/6)$& $1/2$ &$U^c(\bar{3},1,-2/3)+D^c(\bar{3},1,1/3)$&\\
            & $L(1,2,1,-1/2)$& $1/2$ &$L(1,2,-1/2)$&\\
            &$L^c(1,1,2,1/2)$& $1/2$ &$N^c(1,1,0)+E^c(1,1,1)$&\\
            \hline
            &$h(1,2,2,0)$& $0$ &$H_u(1,2,1/2)+H_d(1,2,-1/2)$&\\
            &$\Delta(1,1,3,0)$& $q$ &$\Delta^0(1,1,0)+\Delta^+(1,1,1/3)+$&\\
            & & &$\Delta^-(1,1,-1/3)$&\\
            &${\Phi}(1,1,2,1/2)$& $0$ &$N^c_{\Phi}(1,1,0)+E^c_{\Phi}(1,1,1)$&\\
            &$\bar{\Phi}(1,1,2,-1/2)$& $0$ &$\bar{N}^c_{\Phi}(1,1,0)+\bar{E}^c_{\Phi}(1,1,-1)$&\\
            &$S(1,1,1,0)$& $1$ &$S(1,1,0)$&\\
            &$X(1,1,1,0)$& $1-p$ &$X(1,1,0)$&\\
        \end{tabular}
        \end{ruledtabular}
    \end{table}

The representations  of the various matter and Higgs superfields of the SUSY $G_{3221}$ model as well as their decomposition under SM symmetry and  their charge assignments under $U(1)_{R'}$ symmetry are given in \cref{content}. The superpotential for the left-right symmetric model can be expressed as a combination of renormalizable and nonrenormalizable parts, as
\begin{equation}
	W = W_{\text{r}} + W_{\text{nr}},
\end{equation}
where
\begin{eqnarray}
	W_{\text{r}} &=& W_{\text{S}} +  W_{\text{Y}},  \\
	W_{\text{S}} &=& \kappa S (\Phi \bar{\Phi}-M^2) + \lambda S h^2, \label{ws}\\
	W_{\text{Y}}&=&y_{q}^{ij} Q_i h Q^{c}_{j} + y_{l}^{ij} L_i h L^{c}_{j},
\end{eqnarray}
\noindent and
\begin{eqnarray}\label{wnr}
W_{\text{nr}} &\supset & \beta^{ij} \frac{\Phi^2}{m_P} L^c_i L^c_j + \kappa_p \, S^{p} + \gamma_{n} \, \Delta^{n} \nonumber \\
&+& \lambda'_{n} \, h^{2} \, \Delta^{n}.
\end{eqnarray}
Here,
\begin{eqnarray} \label{kp}
\kappa_p \equiv  \frac{\kappa_X \langle X\rangle}{m_P^{p-2}}, \, \, \gamma_{n} \equiv  \frac{\gamma_{\Delta}\langle X\rangle}{m_P^{n-2}}, \, \, \lambda'_{n} \equiv \frac{\lambda' \langle X\rangle}{m_P^{n}},
\end{eqnarray} 
where $i,\,j=1,2,3$ are the generation indices; $\kappa,\ \lambda,$ $\beta^{ij},\ \kappa_X,\ \gamma_{\Delta}$, and $\lambda'$ are dimensionless couplings; $ m_{P} \simeq 2.4 \times 10^{18}$ GeV is reduced Planck mass; and $y_{q}^{ij}$ and $y_{l}^{ij}$ are the Yukawa couplings. $Q_i$, $Q^{c}_{j}$, $L_i$, and  $L^{c}_{j}$ are matter superfields, and the Higgs superfields include $h$, $\Delta$, $\Phi$, $\bar{\Phi}$, $S$, and $X$. Out of these, $X$ is the gauge-singlet superfield, which is assumed to acquire a nonzero VEV, $\langle X \rangle$, in the hidden sector and cause spontaneous breaking of the $R'$-symmetry. The $R'$-charge of $X$ is $R'[X] = (1 - p)$, ensuring that $R'[XS^p] = 1 = R'[W]$. The exponent $p$ is an integer that controls the level of suppression of the $\slashed{R'}$ terms. The exponent $n$ is an integer that is associated with the $R'$-charge $q$ of $\Delta$, $n = p/q$. As explained later in the section, we require that $n > 2$ so as to ensure a nonzero VEV for $\Delta$.  

Among the various terms that appear in the superpotential given above, $W_{\text{S}}$ and $W_{\text{Y}}$, respectively, contain the superpotential terms involved in inflation and all the Yukawa terms. The first term in $W_{\text{nr}}$ generates masses of RHN. Further, the term with coupling $\gamma_{\Delta}$ (or $\gamma_{n}$) is relevant for generating the first symmetry breaking scale as discussed below. Lastly, the terms involving the couplings $\lambda$ (second term in $W_{\text{S}}$) and $\lambda'$ (fourth term in $W_{\text{nr}}$ as written in terms of $\lambda'_{n}$) are $\mu$-terms generating the electroweak Higgs mass. However, as explained in \cref{muproblem} below, the $\lambda'$ term is significantly suppressed 
compared to the $\lambda$ term
.

The scalar component of the gauge singlet superfield $S$ acts as an inflaton, and the $B-L$ conjugate pair of Higgs superfields ($\Phi$, $\bar{\Phi}$) provides the vacuum energy, $\kappa^2 M^4$, for inflation containing the $B-L$ symmetry breaking scale $M$. To broadly study the effect of $R'$-symmetry breaking terms on the predictions of $\mu$-hybrid inflation, we consider $W_{\text{inf}}$ of the general type,
\begin{eqnarray} \label{spotential}
	W_{\text{inf}}  = \kappa S (\Phi \bar{\Phi}-M^2) + \lambda S h^2 +  \kappa_p S^{p}.
\end{eqnarray} 
To impose leading-order suppression for the term containing $\langle X\rangle S^p$ at the nonrenormalizable level, we set $p \geq 3$. The hybrid inflation model with $p=4$, excluding the $\lambda S h^2$ term, has been previously studied in Ref. \cite{Civiletti:2013cra}. 
Using Eq.~\eqref{spotential}, the global SUSY F-term scalar potential can be calculated as:
\begin{eqnarray}\label{vfglobal}
	V^{\text{global}}_{\text{F}} &\equiv& \sum_i \bigg|\frac{\partial W}{\partial z_i}\bigg|^2  = \left|\kappa (\Phi\bar{\Phi}-M^2)  + \lambda h^2 + p \kappa_p S^{p-1} \right|^2 \nonumber \\
    &+&\kappa^2 |S|^2 (|\Phi|^2 + |\bar{\Phi}|^2) + 4 \lambda^2 |S|^2 |h|^2,
\end{eqnarray} 
where $z_i \in \{S, \Phi, \bar{\Phi}, h\}$ and $|h|^2=|H_u|^2+|H_d|^2$, with $H_u,H_d$ being the electroweak Higgs doublet pair. The D-term potential vanishes along the D-flat direction characterized by $\Phi = \bar{\Phi}^{*}$ and $H_u^i= \epsilon_{ij} H_d^{j*}$. 
Here, $\epsilon_{ij}$ is antisymmetric in its indices with $\epsilon_{12}=1$.

The inflationary potential is described by a three-field framework, $(\Phi=\bar{\Phi}^*, h, S)$. The field $S$ has an associated phase, $\theta_S$, whose mass depends on the value of $S$. Initially, assuming $S$ starts from a very large value $(S \gg M)$ to ensure sufficient inflation, the phase $\theta_S$ rapidly settles to zero, which is also its value at the global minimum of the potential.
This effectively makes $S$ a real field. Meanwhile, the potential is independent of the phases of $\Phi$ and $h$, allowing them also to be treated as real fields. Similar to $\theta_S$, the masses of these fields, given, respectively, by $\kappa^2 |S|^2$ and $4 \lambda^2 |S|^2$, from Eq.~(\ref{vfglobal}), depend on $S$. The large value of $S \gg M$ imparts large masses to these fields, causing them to quickly roll toward their local minima at zero. This reduces the system from three fields to a single effective field, where $S$ alone acts as the inflaton.

To estimate the stability of the inflationary track in the above potential, let us first ignore, for simplicity, the contribution of $\slashed{R'}$ term in the superpotential [this means putting $\kappa_p = 0$ in Eq.~(\ref{spotential})]. In this case, the inflationary track with $\Phi = 0 = h$ is a stable minimum for $S > \max(M,\,M/ \sqrt{\gamma})$, with $\gamma \equiv \lambda/\kappa$ \cite{Dvali:1997uq}. For the $S-\Phi$ system with $h=0$, the critical value of $S$ for stable minimum is $S_c=M$. For the $S-h$ system with $\Phi =0$, the stable minimum is $S_c = M/ \sqrt{\gamma}$. The desired symmetry breaking pattern is where $\Phi$ destabilizes first and acquires a nonzero VEV, $\langle\Phi\rangle = M$, while the $h$ field remains stabilized at the origin. This is achieved by taking $\gamma > 1 $ or $\lambda > \kappa$. 

Let us now return to the original case. Restoring the $\slashed {R'}$ term in the superpotential alters the critical value of the $S$ field for stable minima, up to leading order, as follows:
\[S_c \simeq  \Bigg\{ \begin{array}{lcl}
M - \frac{p}{2} \, \frac{\kappa_p}{\kappa}  \, M^{p-2}, & \mbox{for}
& S-\Phi \,\, \text{system} \\  
\sqrt{\frac{\kappa}{\lambda}} M - \frac{p}{2} \, \frac{\kappa_p}{\lambda}  \, (\sqrt{\frac{\kappa}{\lambda}} M)^{p-2}, & \mbox{for} & S-h \,\, \text{system.}
\end{array}\]
The size of the rescaled coupling $\kappa_p$ is a significant factor in governing the value of the correction terms in both cases. From Eq.~(\ref{kp}), it is easier to see that $\kappa_p$ is not only small, but keeps on decreasing for bigger values of $p$ (i.e., $p \geq 4$). This culminates in infinitesimal corrections in the critical value, leaving it virtually unchanged. Thus, in order to preserve the order of the symmetry breaking, the same condition on the values of $\kappa$ and $\lambda$ (i.e., $\lambda > \kappa$) is applicable in this case as well. 
\subsection{Symmetry breaking}\label{GUTorigin}
The breaking of $SU(2)_R \times U(1)_{B-L}$ into $U(1)_Y$ proceeds in two steps: 
\begin{equation}\label{sb}
SU(2)_R \times U(1)_{B-L}   \xrightarrow{\langle\Delta\rangle}   U(1)_R \times U(1)_{B-L}  \xrightarrow{\langle\Phi \rangle} U(1)_Y  
\end{equation}
Initially, before observable inflation, the gauge group $SU(2)_R$ is assumed to be broken to $U(1)_R$ by a VEV, $\left<\Delta\right> = \Delta^0$, along the more precise direction of $\Delta$.
To explain the origin of the symmetry breaking scale, we consider the nonrenormalizable term in the superpotential,
\begin{equation} \label{xdelta}
W_{\text{nr}} \supset \gamma_{n} \Delta^{n} = \frac{\gamma_{\Delta}\langle X\rangle}{m_P^{n-2}} \Delta^{n},
\end{equation}
where the gauge-singlet superfield $X$ has already acquired a nonzero VEV, $\langle X \rangle = (m_P \ m_{3/2})^{1/2}$, in the hidden sector.
Here, $m_{3/2}$ represents the gravitino mass, indicating the low-energy SUSY breaking scale. This term generates the following soft SUSY breaking scalar potential along with the soft SUSY breaking mass-squared term:
\begin{equation}
V_{\text{soft}} \supset n^2 \gamma_{\Delta}^2 m_{3/2} m_P^3 \left(\frac{|\Delta^0|}{m_P}\right)^{2(n-1)} + 2 m_{3/2}^2 |\Delta^0|^2.
\end{equation}
Assuming $m_{3/2}^2>0$ beyond the symmetry breaking scale, a negative mass squared ($m_{3/2}^2<0$) can arise through the running of renormalization group equations (RGEs), as discussed in Refs. \cite{Ahmed:2022thr, Ahmed:2023rky}, particularly for the $\chi SU(5)$ model. 
The order of the events can also be set through this running effect such that the first breaking occurs before the observable inflation. The first symmetry breaking is triggered due to the negative mass squared term leading to:
\begin{eqnarray} \label{ndelta}
    \left< |\Delta^0| \right>\sim m_P \left(\frac{m_{3/2}}{m_P}\right)^{\frac{1}{2(n-2)}},
\end{eqnarray}
which corresponds to $ \sim 6 \times 10^{14}$~GeV, $ 10^{16}$~GeV,  $2 \times 10^{16}$~GeV, and $5 \times 10^{16}$~GeV, for $n = 4, \, 5, \, 6, \, 7$, respectively, assuming $m_{3/2} \sim 10$~TeV and $\gamma_{\Delta} \sim 1$. Intriguingly, our model can accommodate the typical value of the symmetry breaking scale $\sim 10^{16}$~GeV with $n=5, \,6$ in comparison to the supersymmetric $SU(3)_c\times SU(3)_L\times SU(3)_R$ model \cite{Dvali_1994} where a relatively higher value ($ \sim 3 \times 10^{17}$~GeV with $n=12$) is predicted. Further note that, as described later in \cref{numerics}, the metastable cosmic strings formation also requires $\left< |\Delta^0| \right> \sim 10^{16}$~GeV consistent with both the LIGO-Virgo-KAGRA (LVK) constraints and the NANOGrav 15-year data.

To keep the $SU(2)_R$ breaking scale of order $10^{16}$ GeV, we limit ourselves to $n = 5, \, 6$  along with $p = 3$. This corresponds to $q$ equal to either $3/5$ or $1/2$. In the former case, we can assume that the $R'$-symmetry is spontaneously broken into $Z^{R'}_{10} = Z^{R'}_5 \times Z^{R'}_2$ with the $X$ field acquiring a VEV in the hidden sector. This leaves the $\Delta$ field with an effective $R'$ charge equal to $1/5$. After $\Delta$ acquires a VEV, a $Z^{R'}_2$ symmetry survives and plays the role of matter parity.  As we show later, the $\Delta$ VEV achieved in this scenario is suitable for forming metastable cosmic strings relevant to NANOGrav (\cref{strings}). In such a situation, the $\Delta$ VEV breaks the $Z^{R'}_5$ symmetry, forming domain walls, which are inflated away.

On the other hand, in the latter case ($q=1/2$), the VEV of $\Delta$ breaks the $Z^{R'}_2$ symmetry, hence matter parity, which then has to be imposed by hand to ensure a stable electrically neutral lightest supersymmetric particle as a DM candidate. Thus, the first option with $q=3/5$ is a more attractive choice.

The nonrenormalizable superpotential term given in Eq.~(\ref{xdelta}) can be used to estimate the fermionic masses of $\Delta^+$ and $\Delta^-$ in $\Delta$. Using the $\Delta$ VEV and $\gamma_{\Delta} \sim 1$ as before, we determine that these components have equal masses of order $m_{3/2}$, namely, $\sim [2/(n-1)]^{1/2} m_{3/2}$.

The breaking $SU(2)_R \rightarrow U(1)_R$ produces monopoles, which are quickly diluted away by inflation driven by the $S$ field, as discussed later. After inflation, the second symmetry breaking occurs, where $\Phi$ attains a VEV along $N^c_{\Phi}$ with $h$ remaining at zero. This leads to the formation of metastable cosmic strings \cite{Buchmuller:2021dtt, buchmuller2024metastable}.
\subsection{MSSM $\mathbf{\mu}$-term}\label{muproblem}
The global SUSY minimum is given by
\begin{equation}\label{gsm}
	\langle S\rangle=0 ,  \,\,\,\,\,\, \langle\Phi\bar{\Phi}\rangle=\left<N^c_{\Phi}\bar{N}^c_{\Phi}\right>  =M^2, \,\,\,\,\,\, \langle h\rangle = 0.
\end{equation}
However, due to the effects of soft SUSY breaking terms, the $S$ field acquires a nonzero VEV, $\langle S\rangle \sim m_{3/2}/\kappa$, which generates an effective $\mu$-term, $\mu H_u H_d$, with $\mu \sim (\lambda/\kappa)m_{3/2}$ \cite{Dvali:1997uq,King:1997ia}. Additionally, the following nonrenormalizable term in the superpotential,
\begin{eqnarray}
 W_{\text{nr}} \supset \lambda'_{n} \ h^{2} \Delta^{n} = \frac{\lambda' \langle X\rangle}{m_P^{n}} h^{2} \Delta^{n},
\end{eqnarray}
can also contribute to the $\mu$-term, yielding [using Eq.~(\ref{ndelta})]:
\begin{eqnarray}
    \mu &=&\lambda' (m_P\, m_{3/2})^{1/2}\left(\frac{\Delta^0}{m_p}\right)^{n}\nonumber \\
    &\sim & \lambda' \left(\frac{m_P}{m_{3/2}}\right)^{1/2} \left(\frac{m_{3/2}}{m_P}\right)^{\frac{n}{2(n-2)}}m_{3/2}.
\end{eqnarray}
Comparing the magnitude of these two $\mu$-terms indicates that the latter is significantly suppressed relative to the former, assuming all couplings are of order $1$ and $m_{3/2} \sim 10$ TeV. This suppression starts at roughly a factor of $10^{-7}$ for $n=5$ but reduces for higher values of $n$.
\section{\texorpdfstring{ Minimal $\mathbf{\mu}$}{mu}-Hybrid  Inflation} \label{inflation}
The global SUSY scalar potential in Eq.~(\ref{vfglobal}) remains flat along the valley ($\Phi = 0 = h$), with $V = \kappa^{2} M^4$, making it unsuitable for slow-roll inflation. Nonetheless, several generic contributions to the scalar potential can provide the necessary slope for realistic slow-roll inflation.
First, due to the nonzero vacuum energy term, $\kappa^{2} M^4$, SUSY is broken, which causes a mass splitting between the fermionic and bosonic components of the relevant superfields, resulting in radiative corrections to the scalar potential \cite{Dvali:1994ms}. Second, supergravity (SUGRA) corrections provide another significant contribution \cite{Linde:1997sj}. Lastly, SUSY breaking in the hidden sector also introduces soft SUSY breaking terms in the scalar potential \cite{Buchmuller:2000zm,Senoguz:2004vu,Rehman:2009nq}.

The one-loop radiative corrections can be calculated using the Coleman-Weinberg formula \cite{PhysRevD.7.1888},
\begin{equation} 
\Delta V_{1-\text{loop}}(x)	=  \frac{(\kappa M)^4}{8 \pi^2} \mathcal{N} F(x) + \frac{(\lambda \kappa M^2)^2}{4 \pi^2} F(\sqrt{\gamma}x),
\end{equation} 
where $x \equiv S/M$, $\gamma = \lambda/\kappa$, $\mathcal{N} = 2$ is the dimensionality of the representation of the superfields $\Phi$ and $\bar{\Phi}$ and the loop function $F(x)$ is given by
\begin{eqnarray}
F(x) &=&\frac{1}{4}\bigg((x^4 + 1)\ln{\frac{x^4 - 1}{x^4}} + 2 x^2 \ln{\frac{x^2 + 1}{x^2 - 1}} + \notag \\
 && 2 \ln{\frac{\kappa^2 M^2 x^2 }{Q^2}} -3\bigg),
\end{eqnarray}
with $Q$ being the renormalization scale. The term containing the function $F(\sqrt{\gamma}x)$ arises from the $\lambda S h^2 $ interaction and is dominant for $\lambda > \kappa$.
 
For the SUGRA corrections, we consider the following for the F-term scalar potential,
\begin{equation} \label{SUGRA}
V_{\text{F}} =  e^{K/m_{P}^{2}} \left(K_{ij}^{-1} D_{z_{i}} W D_{z^{\ast}_{j}} W^\ast - \frac{3}{m_{P}^{2}}\vert W\vert ^2\right), 
\end{equation}
where $z_i \in \{S, \Phi, \bar{\Phi}, \ldots\}$ represent the bosonic components of the superfields and
 \begin{align}
&K_{ij} \equiv  \frac{\partial^2 K}{\partial z_{i} \partial z^{\ast}_{j}}, \ \
	D_{z_{i}} W \equiv \frac{\partial W}{\partial z_{i}}+\frac{1}{m_{P}^{2}} \frac{\partial K}{\partial z_{i}} W, \notag\\
&D_{z^{\ast}_{i}} W^\ast = \left(D_{z_{i}} W\right)^\ast.
\end{align}
Regarding the K\"{a}hler potential ($K$), we consider its minimal (canonical) form:
\begin{eqnarray}\label{mkahler}
	K & = & |S|^{2} + |\Phi|^{2}+ |\bar{\Phi}|^{2} + |h|^{2}.
\end{eqnarray}
Using the superpotential of Eq.~(\ref{spotential}) as well as $K$ given above and the SUGRA corrections, the inflationary potential is given by
\begin{eqnarray}\label{vfm}
V_{\text{F}}(S)  & = & V_0 \bigg[ 1 + \frac{S^4}{2\,m_{P}^{4}} + \frac{\kappa_p S^{p-2}}{2 \kappa^2 M^4 m_{P}^4} \bigg\{\kappa_p S^p\bigg(2 \,p^2 \, m_{P}^4 + \notag \\
&& \alpha_1 m_{P}^2 S^2 + \alpha_2 S^4\bigg) - 2 \kappa M^2 S \bigg(2 \, p \, m_{P}^4 + \notag \\
&& \alpha_3 m_{P}^2 S^2 + \alpha_4 S^4\bigg) \bigg\}\bigg],
\end{eqnarray} 
with $V_0  \equiv  \kappa^{2} M^4$, and
\begin{eqnarray}
\alpha_1 = 2(p-1)(p+3) &,& \, \, \alpha_2 = p(p+4)-4 ,\notag \\
\alpha_3 = 4(p-1) &,& \, \, \alpha_4 = 3p-2.
\end{eqnarray}
Finally, the soft SUSY terms are written as \cite{Senoguz:2004vu,Rehman:2009nq}
\begin{equation} 
	V_{\text{\text{soft}}}(x)	= a  \kappa \ m_{3/2}  M^3 x + m_{3/2}^{2} M^2 x^2,
\end{equation}
with
\begin{equation} 
	a = 2 \  \vert 2 - A \vert \cos[\arg S + \arg (2-A)].
\end{equation}
With suitable initial conditions, the phase $\arg S$ can be stabilized before observable inflation \cite{Senoguz:2004vu,Buchmuller:2014epa}. Therefore, we treat $a$ as a constant of order unity.

Including the three types of corrections, the scalar potential, $V $, takes the form along the D-flat direction,
\begin{eqnarray} \label{vx}
V(x)	&\approx &  \ V_{\text{F}} + \Delta V_{1-\text{loop}} + V_{\text{soft}}  \nonumber \\
 &=& V_0 \, \bigg[1+\frac{a \ m_{3/2} \ m_{P}}{\kappa M^2} \left(\frac{M x}{m_{P}}\right)+\frac{m_{3/2}^2 \ m_{P}^2}{\kappa^2 M^4}  \times \notag \\
&& \left(\frac{M x}{m_{P}}\right)^2 + \frac{1}{2}  \left(\frac{M x}{m_{P}}\right)^4 + \frac{\kappa ^2 }{4 \pi^2}  F(x) + \frac{\lambda ^2 }{4 \pi^2}  \times \notag \\
&& F(\sqrt{\gamma}x) + \delta V_p(x) \bigg],
\end{eqnarray}
where $\delta V_p(x)$ includes terms originating from the $R'$-symmetry breaking term $\kappa_p S^p$ in $W_{\text{inf}}$ and is given by
\begin{eqnarray} 
\delta V_p(x)  & = & \frac{\kappa_p m_{P}^{p-2}}{2 \kappa^2 M^4} \left(\frac{Mx}{m_{P}}\right)^{p-2} \bigg[\kappa_p m_{P}^{p} \left(\frac{Mx}{m_{P}}\right)^{p}\bigg\{2 \, p^2  \notag \\
&& + \,\alpha_1  \left(\frac{M x}{m_{P}}\right)^2 + \alpha_2 \left(\frac{M x}{m_{P}}\right)^4\bigg\} - \, 2 \kappa M^2  \times \notag \\
&& m_{P} \, \left(\frac{M x}{m_{P}}\right) \bigg\{2 \, p  + \alpha_3  \left(\frac{M x}{m_{P}}\right)^2 \notag \\
&& + \, \alpha_4 \left(\frac{M x}{m_{P}}\right)^4 \bigg\} \bigg].
\end{eqnarray}
For minimal canonical $K$ without the $\slashed{R'}$ term in $W_{\text{inf}}$, the linear soft SUSY term becomes important for smaller values of $\kappa < 10^{-3}$ with $m_{3/2} \sim 10$~TeV.  In this regime, the SUGRA corrections are negligible, and the radiative correction remains relevant over a wider range of $\kappa$ values \cite{urRehman:2006hu}.
For the SUGRA corrections, some terms appearing in $\delta V_p (x)$ can make significant contributions in the small $\kappa$ regime. It is easy to deduce that terms with lower powers of $(M x/m_P)$ generally dominate. In this context, we can reasonably assume, as confirmed by subsequent numerical analysis, that  $\kappa$  and $\kappa_p $ are much smaller than unity. Therefore, the lower-order terms in the second half of $\delta V_{p}(x)$, which carry an overall negative sign, play a critical role.
\subsection{Slow-roll inflationary parameters}
In the slow-roll approximation, the standard inflationary observables, i.e., the scalar spectral index, $n_s$, the tensor-to-scalar ratio, $r$, and the running of the scalar spectral index, $\alpha_{s}\equiv dn_s / d \ln k$, are written to leading order as
\begin{align} \label{rns}
n_s&\simeq 1-6\,\epsilon+2\,\eta\,\,\,, \,\,\, r\simeq 16\,\epsilon, \notag\\
\alpha_{s} &\simeq 16\,\epsilon\,\eta
-24\,\epsilon^2 - 2\,\zeta^2.
\end{align}
Here, $\epsilon$, $\eta$ and $\zeta$ are the so-called slow-roll parameters, defined in terms of potential $V$ as
\begin{align}
\epsilon &= \frac{1}{2}\left( \frac{m_P}{M}\right)^2
\left( \frac{V'}{V}\right)^2, \,\,\,
\eta = \left( \frac{m_P}{M}\right)^2
\left( \frac{V''}{V} \right), \notag\\
\zeta^2 &= \left( \frac{m_P}{M}\right)^4
\left( \frac{V' V'''}{V^2}\right),
\label{slowroll}
\end{align}
with all derivatives taken with respect to $x$ and $V \equiv V(x)$. The amplitude of the scalar power spectrum, $A_s$, is given by
\begin{align}
A_{s}(k_0) = \frac{1}{24\,\pi^2\,\epsilon}
\left( \frac{V}{m_P^4}\right)\bigg|_{x=x_0},
\end{align}
where $x_0 \equiv x(k_0)$ is the field value at the pivot scale $k_0 = 0.05$ Mpc$^{-1}$ and $A_s(k_0) = ¼ 2.137 \times 10^{-9}$ as measured by Planck \cite{Planck:2018vyg, Planck:2018jri}. The amount of observable inflation is characterized by the number of $e$-folds $N_0$ before the end of inflation and is determined by
\begin{align}\label{Ngen}
N_0 = 2\left( \frac{M}{m_P}\right) ^{2}\int_{x_e}^{x_{0}}\left( \frac{V}{%
	V'}\right) dx,
\end{align}
where $x_e$ is the field value at the end of inflation. For numerical analysis, the value of $x_e$ is fixed either by the breakdown of the slow-roll approximation ($\eta(x_e)=-1$), or by a ``waterfall" destabilization occurring at $x_e = 1$. 
\section{Reheating} \label{reheating}
As the inflationary phase ends, the inflaton system enters the reheating phase, with damped field oscillations around the minimum and eventual thermalization of the Universe by the decaying fields. The inflaton system consists of two complex scalar fields, denoted here as $s$ and $\theta=\left(\delta \phi + \delta \bar{\phi}\right)/\sqrt{2}$ (see Refs. \cite{Lazarides1997,Afzal:2022vjx}, for example) that have a common mass $m_\text{inf} \simeq \sqrt{2}\kappa M$. The dominant decay width, through the direct coupling $\lambda S h^2$, into a pair of Higgsinos ($\widetilde H_u$, $\widetilde H_d$) and Higgses ($ H_u$, $ H_d$), is given by \cite{Lazarides1997}
\begin{equation}\label{gamma}
\Gamma_h =\Gamma(\theta \rightarrow H_u H_d) = \Gamma(s \rightarrow \widetilde H_u\widetilde H_d)=\frac{\lambda ^2 }{8 \pi }m_{\text{inf}}.
\end{equation}
The nonrenormalizable superpotential couplings, $(\beta^{ij} \Phi^2 L^c_i L^c_j)/m_P$, open up another decay channel \cite{Lazarides1997} resulting in a pair of RHN ($N^c$) and sneutrinos ($\widetilde{N}^c$), respectively, with equal decay width $\Gamma_N$ given as
\begin{align}
\Gamma_N& = \Gamma (\theta \rightarrow N^cN^c )= \Gamma(s \rightarrow \widetilde{N}^c\widetilde{N}^c ) \notag\\
&=\dfrac{m_\text{inf}}{8\pi}\left( \frac{M_N}{M} \right)^2\left(1-\dfrac{4M_{N}^{2}}{m_\text{inf}^{2}}\right)^{1/2},
\end{align}
with the condition that only the lightest RHN with mass $M_{N}$ satisfies the kinematic bound, $m_\text{inf} > 2 M_{N}$. The  reheat temperature $T_R$ is defined in terms of the inflaton decay width $\Gamma_\text{inf}$ as

\begin{equation}\label{tr}
T_R=\left(\dfrac{90}{\pi^{2}g_{*}}\right)^{1/4}\left(\Gamma_\text{inf} \, m_{P}\right)^{1/2},
\end{equation}
with
\begin{equation}
\Gamma_\text{inf} = \Gamma_h + \Gamma_N,
\end{equation}
where $g_{*}= 228.75$ for the MSSM. Assuming a standard thermal history, the number of $e$-folds, $N_{0}$, can be written in terms of $T_R$ as \cite{Kolb1990, Liddle:2003as}
\begin{align}\label{efolds}
N_0=53+\dfrac{1}{3}\ln\left[\dfrac{T_R}{10^9 \text{ GeV}}\right]+\dfrac{2}{3}\ln\left[\dfrac{\sqrt{\kappa}\,M}{10^{15}\text{ GeV}}\right].
\end{align}

The $\Gamma_N$ channel, although typically suppressed so that $\Gamma_\text{inf} \simeq \Gamma_h $,  plays a crucial role in implementing successful leptogenesis, which is further discussed in \cref{leptogenesis}. 
\section{Metastable Cosmic Strings} \label{strings}
\begin{figure}
\includegraphics[width=0.49\textwidth]{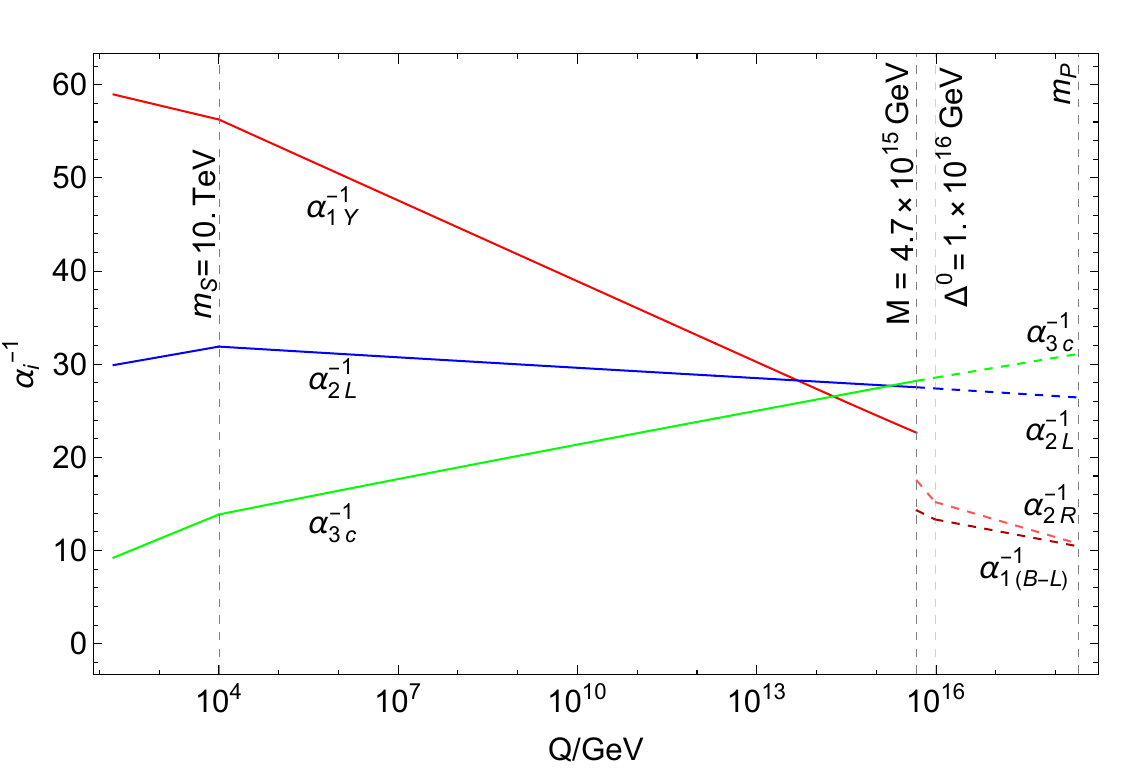}
\caption{\label{gce}Evolution of the two-loop RGEs for gauge couplings in the gauge groups $G_{3221}$, $G_{3211}$, and $G_{\text{SM}}$, assuming all SUSY particles are degenerate at a common mass scale $m_S$.
} 
\end{figure}

The symmetry breaking pattern in Eq.~(\ref{sb}) yields metastable cosmic strings in the usual way. The first breaking generates monopoles that are inflated away. The second breaking produces strings, which can decay by quantum tunneling into string segments connecting monopole-antimonopole pairs. The decay rate per unit string length is given as \cite{Preskill_1993, Leblond_2009, Monin_2008},
\begin{align}
\Gamma_\text{d}&=\frac{\mu_\text{cs}}{2\pi} e^{-\pi \kappa_\text{m}}, \text{  with  } \kappa_\text{m} = \frac{M_\text{m}^2}{\mu_\text{cs}},
\end{align}
where $\mu_\text{cs}$ is the mass per unit length of the string or the string tension, and $M_\text{m}$ represents the monopole mass. The decay time is determined by
\begin{align}
t_\text{cs}=\Gamma^{-1/2}_\text{d}.    
\end{align}
The exponential suppression with $\kappa_\text{m}$ as an exponent in $\Gamma_\text{d}$ suggests that the metastable cosmic string network mimics a stable network for $\sqrt{\kappa_\text{m}}\gg 10$. For $M_\text{m}^2 > \mu_\text{cs}$, the metastable cosmic strings are generally assumed to be effectively stable. Following the approach adopted in Ref. \cite{Buchmuller:2021dtt}, the monopole mass and string tension can be written as
\begin{align}
M_\text{m}& \simeq \frac{4\pi}{g_{2R}} \sqrt{2 v_\Delta^2 + M^2}  \simeq \frac{4\pi}{g_{2R}} \sqrt{2} \, v_\Delta, \notag\\
\mu_\text{cs}& \simeq 4 \pi M^2,
\end{align}
where $v_\Delta$ and $M$, respectively, represent the $SU(2)_{R}$ and $U(1)_R \times U(1)_{B-L}$ symmetry breaking scales, i.e., the formation scales of monopoles and cosmic strings, and $g_{2R}$ is the gauge coupling of the symmetry responsible for monopole generation. We estimate that
\begin{align}
\kappa_\text{m} \sim \frac{8\pi}{g_{2R}^2} \left( \frac{v_\Delta}{M} \right)^2.   \label{eq:alpha} 
\end{align}
For $g_{2R}$ of order unity, the observed NANOGrav signal can be explained by metastable cosmic strings with $\sqrt{\kappa_m} \sim 8$, implying $v_\Delta \sim 2 M$. \cref{gce} illustrates the evolution of the two-loop RGEs of gauge couplings associated with the gauge groups $G_{3221}$, $G_{3211} \equiv SU(3)_c\times SU(2)_L\times U(1)_R\times U(1)_{B-L}$, and $G_{\text{SM}}$.

\section{Viability of Gravitino Dark Matter} \label{dm}
In discussing the viability of gravitino dark matter, we closely follow previous studies that extensively analyzed stable, long-lived unstable, and short-lived unstable gravitinos \cite{Okada:2015vka,Rehman:2017gkm,Lazarides_2021,Ahmed:2021dvo, Afzal:2022vjx, Ahmed:2022wed}. However, we focus on exploring the possibility of a stable gravitino as a cold dark matter candidate within the parameter space defined by our model. This analysis primarily calculates the relic abundance of thermally produced stable gravitinos based on the model parameters.
An approximate but useful relation for determining the relic abundance of stable gravitinos in terms of the reheat temperature, $T_R$, and the gluino mass, $m_{\tilde{g}}$, is expressed as \cite{Bolz:2000fu}
\begin{align}\label{dmab}
\Omega_{3/2}h^2 \simeq 0.08\left(\dfrac{T_R}{10^{10}~\text{GeV}}\right)\left(\dfrac{m_{3/2}}{1~ \text{TeV}}\right)\left(1+\dfrac{m_{\tilde{g}}^2}{3m^2_{3/2}}\right),
\end{align}
where $\Omega_{3/2}=\rho_{3/2}/\rho_{c}$, $h$ is the present Hubble parameter in units of $100$ km $\text{sec}^{-1}$ $\text{Mpc}^{-1}$, $\rho_{3/2}$ is the gravitino energy density, and $\rho_{c}$ denotes the critical energy density of the present-day Universe. This expression is approximate, as it includes only the dominant QCD contributions to the gravitino production rate. The full form of the equation, which incorporates analogous contributions from the electroweak sector, is provided in Refs. \cite{Pradler_2007,PhysRevD.75.075011} and later in Refs. \cite{Eberl_2021, Eberl_2025}. However, for this analysis, the electroweak contributions are ignored. Our primary objective is to determine whether, alongside other findings, the model predicts a dark matter abundance consistent with the result from the Planck satellite, which requires $\Omega_{3/2}h^2\sim 0.12$ \cite{Planck:2018jri}. For that, we apply the LHC constraint on the gluino mass, $m_{\tilde{g}}> 2.2$~TeV \cite{Vami:2019kh}.
\section{Numerical Analysis}\label{numerics}
Having confined this study to the case of a minimal K\"ahler potential, the next critical step in numerically analyzing the predictions of the model is determining realistic values for the parameter $p$, which defines the form of the inflationary potential and thereby influences the inflationary observables. For reasons detailed in \cref{model}, we focus on the case $p=3$ in this analysis. The additional assumptions, approximations, and constraints applied to other free parameters of the model are summarized as follows.

The gravitino overproduction bound on the reheat temperature \cite{KHLOPOV1984265, Ellis:1984eq, PhysRevD.78.065011, PhysRevD.97.023502}, i.e., $T_R  \lesssim 2 \times 10^9$ GeV corresponds to a gravitino mass of $m_{3/2} \sim 10$ TeV. We choose this value to keep the MSSM spectrum within the reach of the LHC and Future Circular Collider \cite{FCC:2018vvp}. From the discussion in \cref{model}, this model constrains us to take $\gamma > 1$. For a benchmark point shown in \Cref{benchmark}, we set $\gamma = 2$. However, to gain a broader understanding of the impact of $\gamma$, we also extend this analysis to other $\gamma$ values allowed by the $T_R$ bounds, as explained in \cref{leptogenesis}. This extension is relevant, as compared to previous work that focused on a single value of $\gamma$ without $\slashed{R'}$ terms (see, for instance, Ref. \cite{Afzal:2022vjx}).

\begin{table}[t]
\caption{\label{benchmark} Benchmark point with $p=3$ in Eq.~(\ref{vx}) or $\kappa_p S^p = (\kappa_X \langle X\rangle S^3)/m_P$ in Eq.~(\ref{spotential}), and $\gamma = 2$.}
\begin{ruledtabular}
\begin{tabular}{cccc}
$\kappa$&$1.6\times10^{-5}$ &  $N_0$&$50.48$\\
\hline
$\lambda$&$3.2\times10^{-5}$ &  $n_s$&$0.965$\\
\hline
$\kappa_X$&$2.8\times10^{-6}$ & $r$&$10^{-13}$\\
\hline
$M $&$4.7\times10^{15}$ GeV  &  $T_R$&$1.5\times10^9\,\text{GeV}$ \\
\hline
$M_N$&$5.4\times10^{10}\,\text{GeV}$ & $G\mu_\text{cs}$ & $9.7\times10^{-8}$\\
\hline
$m_{3/2}$&$10\,\text{TeV}$ & $\Omega_{3/2}h^2$&$0.12$\\
\end{tabular}
\end{ruledtabular}
\end{table}

\begin{figure} [t]
\includegraphics[width=0.5\textwidth]{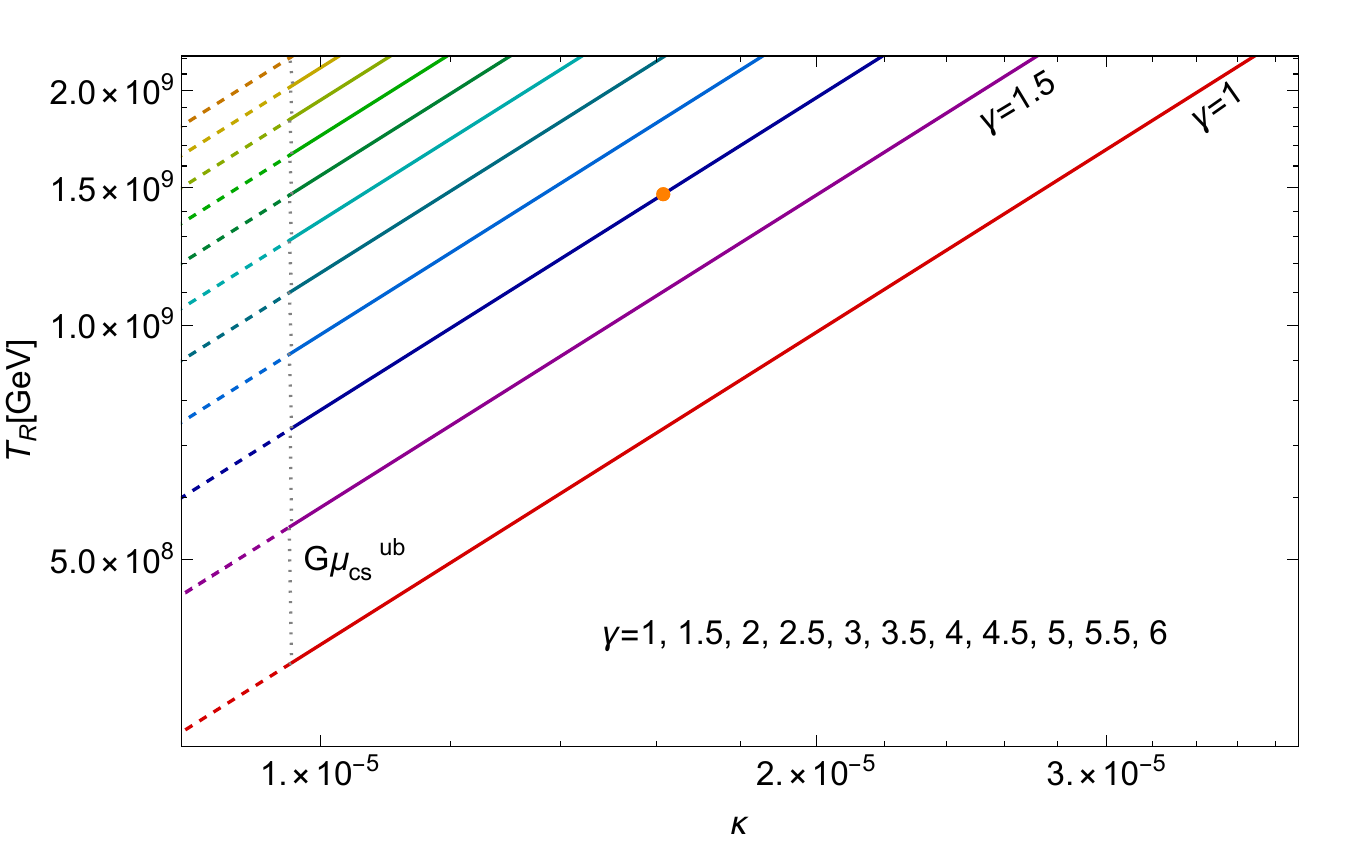}
\includegraphics[width=0.5\textwidth]{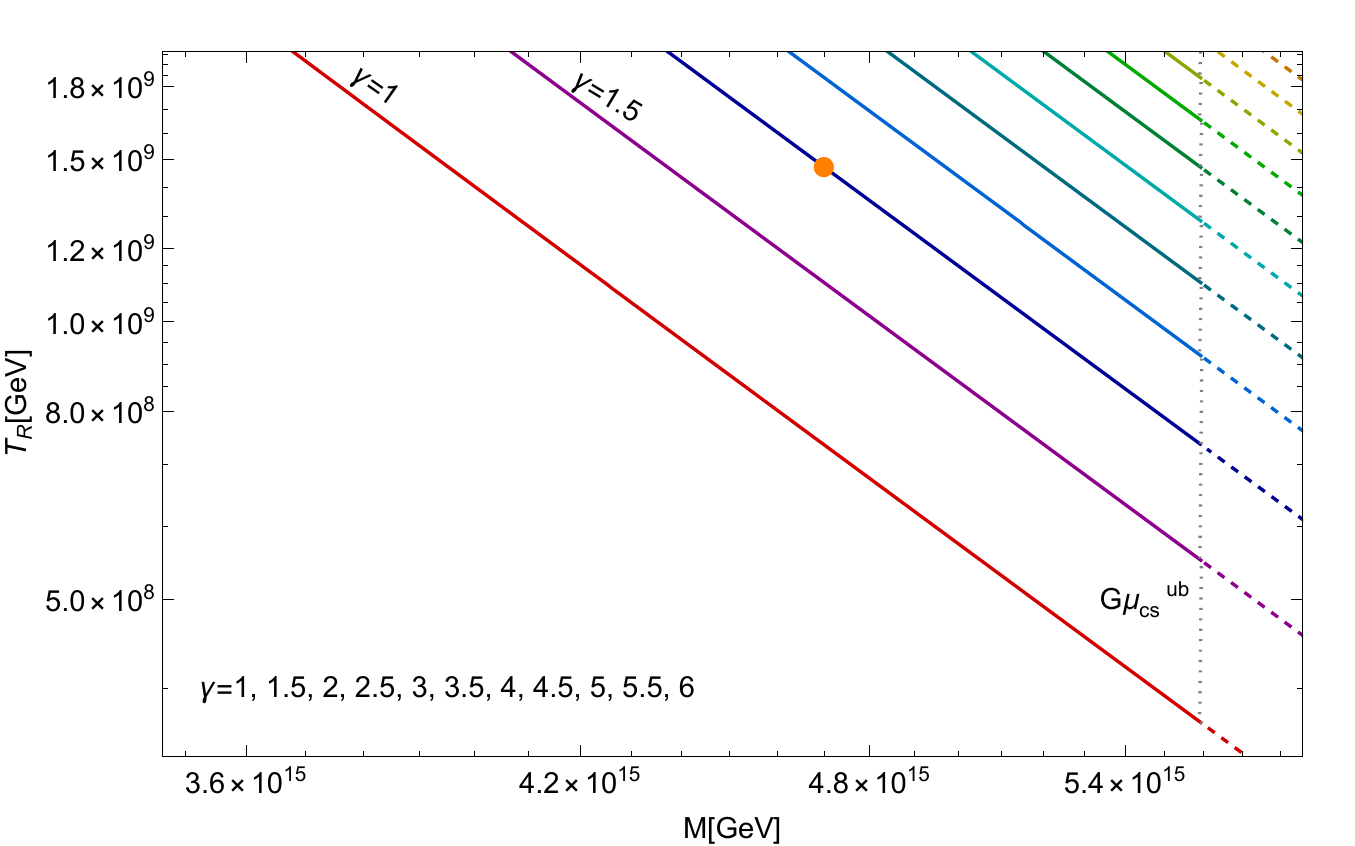}
\caption{\label{plots} Impact of $\gamma$ on the evolution of the reheat temperature $T_R$ as a function of the coupling $\kappa$ (top) and the symmetry breaking scale $M$ (bottom) is shown. Each curve corresponds to a fixed $\gamma$ value, as indicated by different colors. The solid portions of the curves represent $T_R$ and $G\mu_\text{cs}$ values that satisfy both the gravitino overproduction and LVK bounds. The dashed segments, while mathematically consistent, exceed the limit $G\mu_\text{cs} \lesssim 1.3 \times 10^{-7}$, indicated by the dotted vertical line labeled $G\mu_\text{cs}^{\text{ub}}$. The orange dot on the $\gamma=2$ curve denotes the values corresponding to the selected benchmark point.} 
\end{figure}

The renormalization scale $Q$ and the dimensionless constant $a$ are set equal to $M$ and $1$, respectively. This allows flexibility to vary the remaining independent parameters, specifically $\kappa$, $\kappa_X$, $M$, and $M_N$. A key objective of this approach is to derive a model-dependent value of $T_R$, which primarily depends on the parameters $\kappa$ and $M$ through the definitions of $\Gamma_\text{inf}$ and $m_{\text{inf}}$, while having a weaker dependence on $M_N$ due to $\Gamma_N < \Gamma_h$.  For numerical estimation, we begin by adopting the central value of the scalar spectral index $n_s = 0.965$, as reported by the Planck 2018 data \cite{Planck:2018jri}. This setup allows us to determine whether the model can simultaneously predict metastable cosmic strings and gravitino dark matter consistent with experimental observations, while also matching the inflationary CMB observables. This also allows us to implement a satisfactory leptogenesis scenario, as described in \cref{leptogenesis} below. This approach is further strengthened by using Eq.~(\ref{efolds}) to calculate $N_0$ in terms of $T_R$. Additionally, we utilize the Planck  experimental value of the amplitude of the scalar power spectrum, $A_s(k_0) = 2.137 \times 10^{-9}$, at the pivot scale $k_0 = 0.05, \text{Mpc}^{-1}$. We carefully monitor that all approximations and constraints, such as $\Gamma_N < \Gamma_h$ and the relevant kinematic bounds, remain valid throughout. The end of inflation is assumed to occur via the waterfall destabilization condition, i.e., $x_e = 1$.

\Cref{benchmark} presents the values of the various fundamental and derived parameters for the choice $p=3$ and $\gamma=2$, providing a benchmark point that illustrates the numerical calculations based on the aforementioned strategy. This point demonstrates the ability of the model to predict successful and compatible outcomes for leptogenesis, metastable cosmic strings, and gravitino dark matter density. Specifically, for this point, $T_R \simeq 1.5 \times 10^9$ GeV not only produces a sufficient number of $e$-folds to bring $n_s$ to its central value, but is also consistent with the threshold ($\gtrsim 8 \times 10^7$ GeV) associated with the experimental baryon-to-photon ratio, while respecting the kinematic bound. Furthermore, $\Gamma_N$ is suppressed by a factor of $10^{-3}$ compared to $\Gamma_h$ in the model. The predicted string tension $\mu_\text{cs}$ is compatible with the LVK bound given by
\cite{PhysRevD.104.022004, Afzal_2023}
\begin{equation}
G\mu_\text{cs} \lesssim 1.3 \times 10^{-7},
\end{equation} 
where $G$ is the Newton's gravitational constant. Similarly, $\Omega_{3/2}h^2 = 0.12$ aligns with the Planck 2018 data with a gravitino mass of $m_{3/2} = 10$ TeV. It is important to note that the calculated dark matter density is not unique to the choice $\gamma=2$ and can be achieved with other allowed values of $\gamma$ as well, as described below.

The effects of varying $\gamma$ are illustrated in \cref{plots}, where $T_R$ is plotted against both $\kappa$ and $M$. It is clear from the plots that while the bounds from leptogenesis and gravitino overproduction on $T_R$ permit $\gamma$ to reach values as large as 10, the LVK constraint on the maximum allowable value of $G\mu_\text{cs}$ restricts $\gamma$ to the range $1 < \gamma \leq 5$. These bounds help define the viable range for the fundamental model parameters for a given $\gamma$. 
For example, with $\gamma = 2$, the following ranges are obtained: $9.7 \times 10^{-6} \lesssim \kappa \lesssim 2.0 \times 10^{-5}$, $2.3 \times 10^{-6} \lesssim \kappa_X \lesssim 2.9 \times 10^{-6}$, $4.3 \times 10^{15} \lesssim M \text{(GeV)} \lesssim 5.5 \times 10^{15}$, $3.8 \times 10^{10} \lesssim M_N \text{(GeV)} \lesssim 6.3 \times 10^{10}$, and $7.6 \times 10^{10} \lesssim m_{\text{inf}} \text{(GeV)} \lesssim 1.2 \times 10^{11}$.

These already narrow ranges shrink further as $\gamma$ increases, as shown in \cref{plots} for certain parameters. The lowest value of $T_R$ in the allowed range, which drops to $7.5 \times 10^{8}$ GeV for $\gamma = 1.5$, steadily increases with rising $\gamma$, reaching $1.8 \times 10^{9}$ GeV for $\gamma = 5$. This limits the flexibility of other parameters as $\gamma$ increases. The values of $\kappa$, $T_R$, and $M$ corresponding to the benchmark point are highlighted by an orange dot on the $\gamma = 2$ curve in the plots.
\section{Nonthermal Leptogenesis} \label{leptogenesis}
The $\Gamma_N$ channel described in \cref{reheating} also contributes to the observed baryon asymmetry through the sphaleron process \cite{Kuzmin:1985mm,Fukugita:1986hr,Khlebnikov:1988sr}. The suppression of the washout factor of lepton asymmetry can be achieved if $M_N \gg T_R$ is assumed. The observed baryon asymmetry can be estimated in terms of the lepton asymmetry factor $\varepsilon_L$:
\begin{align}\label{bphr}
\frac{n_{B}}{n_{\gamma}}\simeq -1.84 \,\varepsilon_L  \frac{\Gamma_N}{\Gamma_\text{inf}}\frac{T_R}{m_\text{inf}} \delta_{\text{eff}},
\end{align} 
where $\delta_{\text{eff}}$ is the $CP$-violating phase factor, $\Gamma_\text{inf} \simeq \Gamma_h$, and
\begin{eqnarray}
(-\varepsilon_L)  \simeq \frac{3}{8\pi}  \frac{\sqrt{\Delta m_{31}^{2}} M_N}{\langle H_{u}\rangle^{2}},
\end{eqnarray} 
assuming hierarchical neutrino masses \cite{Rehman:2018gnr}. Here, the atmospheric neutrino mass squared difference is $\Delta m_{31}^{2}\approx 2.6 \times 10^{-3}$ eV$^{2} $  and $\langle H_{u}\rangle \simeq 174$ GeV in the large $\tan\beta$ limit. 

For the observed ratio $n_\mathrm{B} /n_\gamma = (6.12 \pm 0.04) \times 10^{-10}$ \cite{ParticleDataGroup:2020ssz}, the constraint on $|\delta_{\text{eff}}|\leq 1$ along with the kinematic bound, $m_\text{inf} \geq 2M_N$, corresponds to the following bound on the reheat temperature \cite{Afzal:2022vjx}: 
\begin{align}\label{lept}
T_R  \gtrsim \gamma^2\left(2 \times 10^7\right) \text{ GeV}.
\end{align} 

Interestingly, this $T_R$ bound for successful leptogenesis can be used to provide a qualitative estimate of the upper limit for the value of $\gamma$, without going into the exact numerical details. If read in conjunction with the similar bound implied by the gravitino overproduction, it turns out that we have some freedom given by $1 < \gamma \lesssim 10$, which, of course, is further refined once the LVK bound is taken into account, as shown above in \cref{numerics}.

 \begin{figure} [t]
\includegraphics[width=0.5\textwidth]{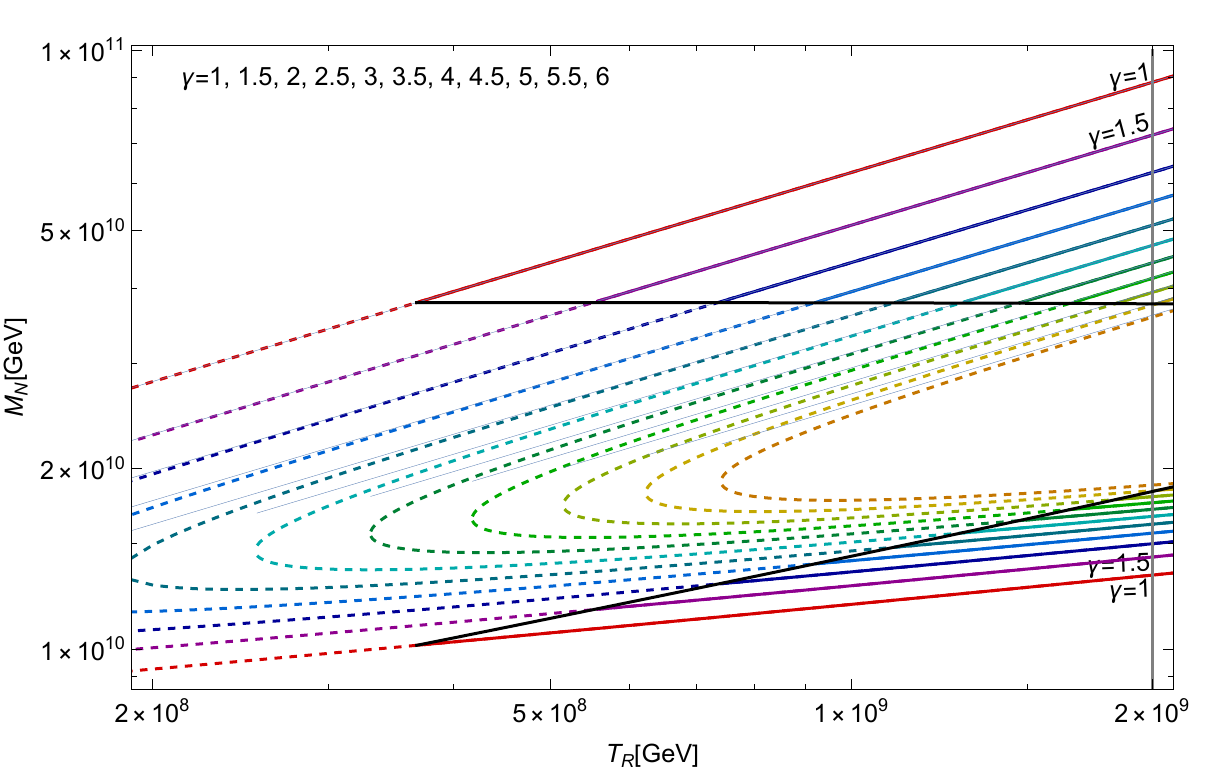}
\caption{\label{plot2} $M_N$ is plotted against $T_R$, illustrating not only the effect of increasing $\gamma$ on the relationship between these parameters but also revealing a degeneracy in the value of $M_N$. As before, the different colors of the curves correspond to fixed values of $\gamma$, as indicated. A vertical line on the right marks the gravitino overproduction bound for $T_R$. The solid gray lines, associated with various $M_N$ vs $T_R$ curves, represent $m_{\text{inf}}/2$ and provide a comparison to the kinematic bound. The solid black lines highlight the slice corresponding to $G\mu_\text{cs}^{\text{ub}}$ for varying $\gamma$. The solid portions of each $M_N$ vs $T_R$ curve refer to $T_R$ and $G\mu_\text{cs}$ values that fall within the bounds for gravitino overproduction and LVK, respectively, while the dashed sections extend beyond the $G\mu_\text{cs}^{\text{ub}}$ threshold.} 
\end{figure}

To quantitatively assess whether the model can also predict consistent leptogenesis, we employ the observed value of $n_\mathrm{B}/n_\gamma$, along with $|\delta_{\text{eff}}| = 1$, which helps in determining the compatible range for $M_N$. One such value is given as part of the benchmark point data shown in \cref{benchmark}. These data indicate a significant suppression of washout effects in nonthermal leptogenesis with $M_N \simeq 34 \, T_R$.

Besides, \cref{plot2}, which essentially presents the relationship between the RHN mass and reheat temperature based on the numerical calculations, indicates a degeneracy in the values of $M_N$. The solid-colored curves in the plot represent compatibility with $G\mu^{\text{ub}}_{\text{cs}}$, while the corresponding dashed curves extend beyond this threshold. This degeneracy stems from the nonlinear dependence of the baryon-to-photon ratio on $M_N$, which is at the least quadratic considering the kinematic bound. A straightforward qualitative analysis of Eq.~(\ref{bphr}), along with the other approximations used, confirms this estimate. However, the consistency of the predictions of the model hinges on the compatibility of $M_N$ solutions with the kinematic bound, which corresponds to the larger of the two $M_N$ values for each $\gamma$, shown as solid gray lines in \cref{plot2}. The black lines in the same plot highlight the data slice for $G\mu^{\text{ub}}_{\text{cs}}$ as $\gamma$ varies. Notably, as $\gamma$ increases, the difference between the two $M_N$ values corresponding to the same $T_R$ value decreases significantly. For instance, at $T_R = 2 \times 10^{9}$ GeV, the approximate order-of-magnitude difference between the $M_N$ values on the $\gamma=1$ curve is reduced to a factor of about 2 for $\gamma=5$.
\section{\texorpdfstring{$\mathbf{G_{3221} \subset G_{422}}$}{3221 to 422} Embedding} \label{422emb}
\begin{figure}[t!]
    \centering
    \includegraphics[width=1.0\linewidth]{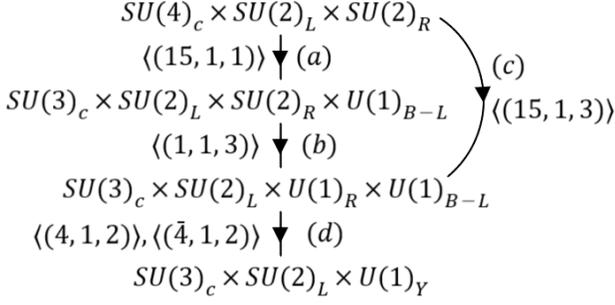}
    \caption{\label{422}Breaking patterns of the gauge symmetry $G_{422}$ down to the SM gauge symmetry $G_{\text{SM}}$.}
\end{figure}

As mentioned in \cref{intro}, the subgroup $G_{3221}$ can be embedded within the larger gauge group $G_{422}$ \cite{PhysRevD.10.275}. The symmetry breaking sequence from $G_{422}$ to $G_{\text{SM}}$ is outlined in \cref{422}, highlighting the different stages of symmetry breaking. Initially, $G_{422}$ breaks down to $G_{3221}$ via the VEV equal to $\langle (15,1,1)\rangle$ (a), followed by the transition of $G_{3221}$ to $G_{3211}$ through $\langle (1,1,3)\rangle$ (b). Alternatively, $G_{422}$ can break directly to $G_{3211}$ via $\langle (15,1,3)\rangle$ (c). In both scenarios, $G_{3211}$ eventually breaks down to $G_{\text{SM}}$ through $\langle (4,1,2)\rangle$ and $\langle (\bar{4},1,2)\rangle$ (d). This study focuses on exploring the $G_{422}$ symmetry breaking, specifically examining pathways (a) and (b).

\begin{table}[t]
    \caption{\label{422content} Decomposition of $G_{422}$ content into $G_{3221}$ representations and their $U(1)_{R'}$ charges.}
    \begin{ruledtabular}
        \begin{tabular}{ccc}
         $\mathbf{G_{422}}$&$\mathbf{q\bigl(U(1)_{R'}\bigr)}$&
        \qquad$  \mathbf{G_{3221}}$ \\
        \hline
         $F_i(4,2,1)$  &$1/2$& $Q_i(3,2,1,1/6)+L_i(1,2,1,-1/2)$ \\
         $F^c_i(\bar{4},1,2)$  &$1/2$&$Q^c_i(\bar{3},1,2,-1/6)+{L}^c_i(1,1,2,1/2)$ \\
        \hline
         $H^c(\bar{4},1,2)$ &$0$&$ \Phi(1,1,2,1/2)+\Phi'(\bar{3},1,2,-1/6)$ \\
         $\bar{H}^c(4,1,2)$ & $0$ &$\bar{\Phi}(1,1,2,-1/2)+\bar{\Phi}'(3,1,2,1/6)$ \\
         $h(1,2,2)$& $0$ & $h(1,2,2,0)$ \\
         $G(6,1,1)$ & $1$ & $g_a(3,1,1,-1/3)+g^c_a(\bar{3},1,1,1/3)$\\
         $S(1,1,1)$ & $1$ & $S(1,1,1,0)$\\
        \end{tabular}
    \end{ruledtabular}
\end{table}

\begin{table}[!]
\caption{\label{addcontent}Decomposition of new superfields, added in $G_{3221}$ and $G_{\text{SM}}$, and their $U(1)_{R'}$ charges. }
    \begin{ruledtabular}
        \begin{tabular}{ccc}
            $\mathbf{G_{3221}}$&
            $\mathbf{q\bigl(U(1)_{R'}\bigr)}$&
            $\mathbf{G_{\text{SM}}}$\\
            \hline
            ${\Phi}' \left( \bar{3}, 1,2, -1/6 \right)$ &  $0$     & $ {d}^c_H\left( \bar{3}, 1, 1/3 \right)$ \\
			& & $ {u}^c_H\left(\bar{3}, 1, -2/3 \right)$\\
			$\bar{\Phi}' \left( 3, 1,2, 1/6 \right)$ &  $0$      & $\bar{d}^c_H\left( 3, 1, -1/3 \right)$ \\
			&  & $\bar{u}^c_H\left(3, 1, 2/3 \right)$\\
             $g_a(3,1,1,-1/3)$& $1$ & $g_a(3,1,-1/3)$  \\
            $g^c_a(\bar{3},1,1,1/3)$&$1$ & $g^c_a(\bar{3},1,1/3)$ \\
            \end{tabular}
    \end{ruledtabular}
\end{table}

The $G_{422}$ representations containing the $G_{3221}$ content are outlined in \cref{422content}, including their respective assignments of $U(1)_{R'}$ charges. Within this framework, the matter content of the MSSM and the superfields for RHN are present in the representations $F_i(4,2,1)$ and $F^c_i(\bar{4},1,2)$ of $G_{422}$. The Higgs sector is distributed among the representations $H^c(4,1,2)$, $\bar{H}^c(\bar{4},1,2)$, and $h(1,2,2)$. To embed $G_{3221}$ in $G_{422}$, we need to add the superfields $\Phi'$ and $\bar{\Phi}'$ in $G_{3221}$ listed earlier in \cref{content}. The decomposition of $\Phi'$ and $\bar{\Phi}'$ under $G_{3211}$ and $G_{\text{SM}}$ along with their corresponding $U(1)_{R'}$ charges is given in \cref{addcontent}. The addition of these superfields results in the appearance of the following nonrenormalizable terms in the superpotential (with generation indices of the superfields suppressed):
\begin{eqnarray}
W_{\text{nr}}^{\text{add}} &\supset & \frac{ \Phi' \Phi}{m_P} (\gamma_3 Q L + \gamma_4 Q^c Q^c)
\nonumber \\
&+&\frac{ \bar{\Phi}' \bar{\Phi} }{m_P}(\gamma_5 Q Q + \gamma_6 Q^c L^c)\nonumber \\
&+& \frac{ \bar{\Phi}' \bar{\Phi}'}{m_P}(\gamma_7 Q L+\gamma_8 Q^c Q^c),
\end{eqnarray}
where $\gamma_3, \gamma_4,\ldots, \gamma_8$ represent dimensionless couplings. These additional terms contribute to proton decay. Thus, the content of our $G_{3221}$ model explains the predictions of proton decay in the $G_{422}$ model \cite{Lazarides:2020bgy}. Integrating out the color triplets, we effectively obtain proton decay operators. For the color triplets $(d_H, \bar{d}_H)$ to acquire mass, $G_{422}$  and $G_{3221}$ must contain a sextet superfield $G \supset g_a+g^c_a$ such that
\begin{eqnarray}
    W_{\text{G}} &\supset & \lambda_{T}\ g_a \Phi \ \Phi' + \lambda_{\bar{T}}\ g^c_a \bar{\Phi}\ \bar{\Phi}',
\end{eqnarray}
where $\lambda_{T}$ and $\lambda_{\bar{T}}$ are dimensionless couplings. The masses of the color  triplets $d^c_{\Phi}$ and $\bar{d}^c_{\Phi}$ are given by,
\begin{eqnarray}
    M_{d^c_{\Phi}}=\lambda_{\bar{T}} M,\qquad M_{\bar{d}^c_{\Phi}}=\lambda_{T} M.
\end{eqnarray}

The decomposition of $G \supset g_a + g_a^c$ and its $U(1)_{R'}$ charges are also listed in \cref{addcontent}. The proton decay predictions for the $G_{3221}$ model are consistent with those previously discussed for the $G_{422}$ model in Ref. \cite{Lazarides:2020bgy}. Another potential embedding within the $G_{422}$ framework, specifically $G_{421} \equiv SU(4)_c \times SU(2)_L \times U(1)_R$, has been explored in Ref. \cite{ahmed2024inflation}. In this scenario, the proton decay predictions fall within the observable range for the next-generation experiments, assuming a color triplet mass of the order of $10^{12}$ GeV.

It is important to emphasize that the $G_{422}$ symmetry is not grand unification since there are three distinct gauge couplings at high scale (see \Cref{gce}). However, unification can be achieved by introducing additional matter fields, allowing the model to be embedded within an $SO(10)$ framework. A similar nontrivial embedding of a left-right symmetric model has been recently explored, for example, in Ref. \cite{Pallis:2024joc}.

Notably, these predictions are derived from LLRR-type proton decay operators, which are characterized by their chirality nonflipping nature. Such operators have been thoroughly studied in other GUTs, including flipped $SU(5)$ \cite{Mehmood:2020irm, Abid:2021jvn}, $SU(5)$ model with the missing doublet mechanism and GUT scale Higgs in the $75$ representation \cite{Mehmood:2023gmm}, and the $R'$-symmetric $SU(5)$ model using the missing doublet mechanism with GUT scale Higgs in the $24$ representation \cite{Ijaz:2023cvc}.

\section{Conclusion}\label{con}
Building on a successful realization of $\mu$-hybrid inflation within the left-right symmetric gauge group $G_{3221}$, we have explored various cosmological implications of the model.  The proposed framework involves a sequential symmetry breaking process, where the breaking of $SU(2)_R \rightarrow U(1)_R$ generates monopoles, followed by the breaking of $U(1)_R \times U(1)_{B-L} \rightarrow U(1)_Y$, which produces cosmic strings. Observable inflation occurs after the first symmetry breaking, with a gauge singlet inflaton originating from a very large value, ensuring sufficient inflation.

Following similar previous studies, we utilize soft SUSY breaking and one-loop radiative corrections to generate the necessary slope for the inflaton to roll. However, a novel aspect of our approach is the inclusion of $\slashed{R'}$ terms in the superpotential to assess their impact on the results. The mathematical framework adopted for this scheme yields values for the reheat temperature $T_R$ and the RHN mass $M_N$, primarily influenced by fundamental parameters such as $\kappa$, $\lambda$, $\kappa_X$, and $M$. Consequently, the dependence of $N_0$, $n_{B}/n_{\gamma}$, and $\Omega_{3/2}h^2$ on $T_R$, as well as that of $G\mu_\text{cs}$ directly on $M$, allows for smooth integration and mutual compatibility of the testable observables predicted by the model.

Numerical evaluation of the model demonstrates its ability to simultaneously implement successful leptogenesis and the metastable cosmic string scenario, and realize gravitino dark matter density, consistent with the observational data. Throughout the analysis, the model adheres to its assumptions and maintains the integrity of its framework, while staying within the bounds on the reheat temperature $T_R$ derived from observational constraints on leptogenesis, gravitino overproduction, and the LVK limit on the string tension. Given a connection between these aspects, we also briefly explore the potential embedding of $G_{3221}$ into $G_{422}$ gauge symmetry, including a discussion of observable proton decay predictions as an addendum.



\FloatBarrier
\bibliographystyle{apsrev4-1}
\bibliography{bibliography}
\end{document}

%% file: authors.tex
\author{Muhammad Nadeem Ahmad
\orcidlink{0009-0002-1072-4873}}
\email{muhnadeem.ahmad@hotmail.com}
\affiliation{Department of Physics, Quaid-i-Azam University, Islamabad 45320, Pakistan}
\author{Maria Mehmood 
\orcidlink{0000-0002-3792-8561}}
\email{mehmood.maria786@gmail.com}
\affiliation{Department of Physics, Quaid-i-Azam University, Islamabad 45320, Pakistan}
\author{Mansoor Ur Rehman
\orcidlink{0000-0002-1780-1571}}
\email{mansoor@qau.edu.pk}
\affiliation{Department of Physics, Quaid-i-Azam University, Islamabad 45320, Pakistan}
\affiliation{Department of Physics, Faculty of Science, Islamic University of Madinah, 42351 Madinah, Saudi Arabia}
\author{Qaisar Shafi}
\affiliation{Bartol Research Institute, Department of Physics and Astronomy,
University of Delaware, Newark, Delaware 19716, USA}